\documentclass[10pt,aps,prd,twocolumn,showpacs,amsmath,amssymb,nofootinbib,eqsecnum,preprintnumbers,superscriptaddress]{revtex4-2}
\usepackage{amsmath,amssymb}
\usepackage{enumitem}  
\usepackage[usenames, dvipsnames]{color} 
\usepackage{graphicx} 
\usepackage{comment}
\usepackage[normalem]{ulem}
\usepackage[utf8]{inputenc}
\usepackage{url}
 \usepackage{xcolor}

\usepackage{hyperref}  

\usepackage{graphicx}
\usepackage{dcolumn}
\usepackage{bm}

\newcommand{\dd}{{\rm{d}}} 

\newcommand{\be}{\begin{equation}}
\newcommand{\ee}{\end{equation}}
\newcommand{\ba}{\begin{eqnarray}}
\newcommand{\ea}{\end{eqnarray}}

\newcommand{\beq}{\begin{equation}}
\newcommand{\eeq}{\end{equation}}
\newcommand{\beqa}{\begin{eqnarray}}
\newcommand{\eeqa}{\end{eqnarray}}

\begin{document}

\title{Investigation of global structure of Kerr-Newman-Bertotti-Robinson spacetime}


\author{Hryhorii Ovcharenko}

\email{hryhorii.ovcharenko@matfyz.cuni.cz}

\affiliation{Institute of Theoretical Physics, Faculty of Mathematics and Physics,
Charles University, Prague, V Hole{\v s}ovi{\v c}k{\' a}ch 2, 180 00 Prague 8, Czech Republic}

\date{\today}


\begin{abstract}
We conduct an investigation of the global structure of Kerr-Newman-BR spacetime and its various subcases such as Schwarzschild-BR spacetime. In this work, we extend the result of our previous studies, where we found that $r\to \infty$ is not the proper conformal infinity. By using the reciprocal coordinate $y=1/r$, we show that the whole family of Kerr-Newman-BR spacetimes can be continuously extended through the surface $r\to \infty$. However, the conformal structure after passing $r\to \infty$ strongly depends on the type of spacetime one is considering (either rotation is present, or not) and on the polar direction one chooses to be fixed (namely, whether we consider generic polar angle $\theta$, or we focus on the behaviour near the equatorial plane $\theta=\pi/2$, or near the poles $\theta=0,\pi$). Physically this extension can be interpreted as kind of ``confinement box,'' created solely by the magnetic field. Finally, we discuss the universality of such a continuous extension.   
\end{abstract}

 \maketitle

\section{Introduction}

Investigation of the global structure is one of the most interesting and challenging parts of the study of new exact spacetimes. The basic understanding of its importance was presented in the pioneering works by Roger Penrose \cite{Penrose1963} and Brandon Carter \cite{Carter1966,Carter1968}. 

The importance of investigating global structure is caused by the fact that understanding the position of the horizons and conformal infinities is required while considering various aspects of the given spacetime, such as gravitational radiation and thermodynamics. In particular, methods of investigation of gravitational radiation are different depending on whether the spacetime is asymptotically flat \cite{Alvares2022}, asymptotically de Sitter \cite{Alvares2022_2}, or anti-de-Sitter \cite{Alvares2026}; see also the recent applications \cite{Podolsky2026_1,Alvares2026_2,Henriques2026}. Concerning thermodynamics, the holographic methods may be applied only if the spacetime is asymptotically AdS \cite{Kubiznak2017,Papadimitrou2005,Abalon2018}. 

However, for a given spacetime, its global structure cannot usually be deduced in the standard coordinates. The significant issue that frequently appears is related to the fact that one is using coordinates that cover only a given coordinate patch, but not the whole manifold. Because of that, it is important to \textit{smoothly extend} the given patch such that at the end one obtains a \textit{geodesically maximal} manifold \cite{Szekeres1960, Kruskal1960}. This task is not always easy, and sometimes it requires complicated analysis. See various examples in \cite{Griffiths2009}.

In particular, recently there appeared a new class of spacetimes \cite{Ovcharenko2025} with an external non-aligned electromagnetic field. From an analysis of one of its special cases, namely the Kerr-BR spacetime \cite{Podolsky2025}, it became clear that this external field is of the Bertotti-Robinson type, so the whole class represents black holes in the external uniform Bertotti-Robinson electromagnetic field. Recently, we were able to explicitly discover the Kerr-Newman-BR spacetime \cite{Ovcharenko2026_2}, identify a genuine Kerr-BR spacetime without electric charge, and investigate its thermodynamics \cite{Kubiznak2026}. This class of spacetimes has already found wide astrophysical applications \cite{Zeng2025,Wang2026,Zhang2026,Li2026,uzbek2026}, but understanding its conformal structure remains an open question. Namely, one could expect that as the Bertotti-Robinson spacetime has $\mathrm{AdS}_2\times S^2$ topology, one would reach the $\mathrm{AdS}$ boundary at large $r$'s. However, this is not the case, as we showed in \cite{Podolsky2025}, and $r\to +\infty$ is not the conformal boundary. Thus, two important questions can be raised because of this: (i) where is the conformal boundary for the whole class of Kerr-Newman-BR spacetimes, (ii) what happens when the observer reaches $r\to \infty$. Also, it is interesting to notice that even though the thermodynamics of this class of spacetime was investigated \cite{Hu2026,Kubiznak2026,Zhang2026_0}, these works did not calculate the corresponding thermodynamic quantities measured on the proper physical boundary. Our work aims to provide an understanding that can be useful for thermodynamic calculations as well.

While this work was in preparation, another paper \cite{Zhou2026} appeared where the authors considered the global structure of the original Kerr-BR spacetime \cite{Podolsky2025}. Even though the technical points of the derivations that we obtained independently, coincide partially, several points were still not covered in \cite{Zhou2026}, namely, the investigation of the Schwarzschild-BR spacetime was not done, the conformal structure in the equatorial plane was not studied, and, most importantly, it is not clear whether the same result holds in the case when the charge of the black hole is non-zero.

To answer these open questions, we start with the easier task in Section II, namely, we investigate the global structure of the Schwarzschild-BR spacetime, where we develop the main ideas of how to continuously extend the spacetimes immersed in magnetic fields. Then, in Section III, we move to the investigation of the new Kerr-Newman-BR spacetime. At the end, in Section IV, we conclude the results of this work and discuss the universal applicability of the continuous extension obtained in this work.

\section{Conformal structure of the Schwarzschild-BR spacetime}

Let us start by reviewing the Schwarzschild-BR spacetime. For this, we, instead of its original form, presented recently in \cite{Podolsky2025}, use the metric form presented in equation (3.13) in \cite{Ovcharenko2026_2}:
\begin{align}
    \dd s^2=\dfrac{1}{\Omega^2}\Big[-\mathcal{Q}\,\dd t^2+\dfrac{\dd r^2}{\mathcal{Q}}+r^2\Big(\dfrac{\dd \theta^2}{P}+P \sin^2\theta\, \dd \varphi^2\Big)\Big],\label{Schw_BR_metr}
\end{align}
where 
\begin{align}
    P&=1+B^2m^2 \cos^2\theta\label{P_Schw_BR}\\
    I&=(1-B^2\,m\,r)^2+B^2r^2,\\
    \mathcal{Q}&=I\Big((1+B^2m^2)-\dfrac{2m}{r}\Big),\label{Q_Schw_BR}\\
    \Omega^2&=I-B^2r^2\Big((1+B^2m^2)-\dfrac{2m}{r}\Big)\cos^2\theta.\label{Om_Schw_BR}
\end{align}

The Weyl scalar $\Psi_2$ in these coordinates takes the form:
\begin{align}
    \Psi_2=\dfrac{m}{r^3}(1-B^2 m r \sin^2\theta)\,\Omega^2.\label{Psi_2_Schw_BR}
\end{align}

The basic issue is that $r\to +\infty$ limit does not lead to the vanishing of the Weyl tensor, and because of that, $r\to +\infty$ does not represent the proper boundary of the Schwarzschild-BR spacetime (unlike the case without the magnetic field $B=0$). 

The global structure of the Schwarzschild-BR spacetime will be understood if we continuously extend the spacetime given by (\ref{Schw_BR_metr})-(\ref{Om_Schw_BR}) across the $r\to+\infty$ boundary and construct a geodesically maximal manifold. According to the definition, a geodesically maximal manifold is bounded either (i) by physical singularities, or (ii) by surfaces at which the affine parameter of geodesically moving particles diverges. Thus, we have to understand the structure of singularities of this spacetime and find surfaces where the affine parameter of the geodesic particles becomes infinite. 

The structure of the singularities can be easily deduced by inspecting (\ref{Psi_2_Schw_BR}). One directly sees that the singularity is placed at $r=0$.The situation with the surface at which the affine parameter of geodesic particles diverges is more complicated. For null particles, this surface is defined by the zeros of the conformal factor $\Omega$ (this can be seen from \cite{Wald1984}, Appendix D. The corresponding argument holds as the metric \ref{Schw_BR_metr} without the conformal factor is regular everywhere except for the singularity). The timelike particles cannot reach this surface at all; see Appendix \ref{append}. Thus, for any type of particles, $\Omega=0$ indeed serves as the proper physical boundary. As we show below, it is easier to investigate the condition $\Omega=0$ in another set of coordinates $x$ and $y$, instead of the coordinates $\theta$ and $r$. The coordinates $x$ and $y$ are defined as:
\begin{align}
    x=\cos\theta,~~~~~~y=\dfrac{1}{r}.
\end{align}

The metric in these coordinates takes the form
\begin{align}
    ds^2=\dfrac{1}{\Omega_y^2}\Big[-\mathcal{Q}_y\dd t^2+\dfrac{\dd y^2}{\mathcal{Q}_y}+\dfrac{\dd x^2}{P_y}+P_y \dd \varphi^2\Big],\label{metr_y}
\end{align}
where 
\begin{align}
    P_y&=(1-x^2)(1+B^2m^2 x^2),\\
    I_y&=(y-B^2m)^2+B^2,\\
    \mathcal{Q}_y&=I_y\Big((1+B^2m^2)-2 m\,y\Big),\\
    \Omega_y^2&=I_y-B^2\Big((1+B^2m^2)-2 m\,y\Big)x^2.\label{Om_y}
\end{align}

In these coordinates, the solution of the equation $\Omega=0=\Omega_y$ takes the form:
\begin{align}
    y_{\pm}=m\,B^2  (1-x^2)\pm B \sqrt{(x^2-1)(1+B^2 m^2 x^2)}.
\end{align}

Let us investigate these solutions. First of all, we notice that in the range of $x$, corresponding to the polar coordinates on the sphere (without the poles), namely for $x\in (-1,1)$, the corresponding roots for $y_{\pm}$ are \textit{complex}. The only non-trivial root appears for $x=\pm 1$ (that corresponds to the poles of the sphere); in this case $y_{\pm}=0$.

\begin{figure}
    \centering
    \includegraphics[width=1\linewidth]{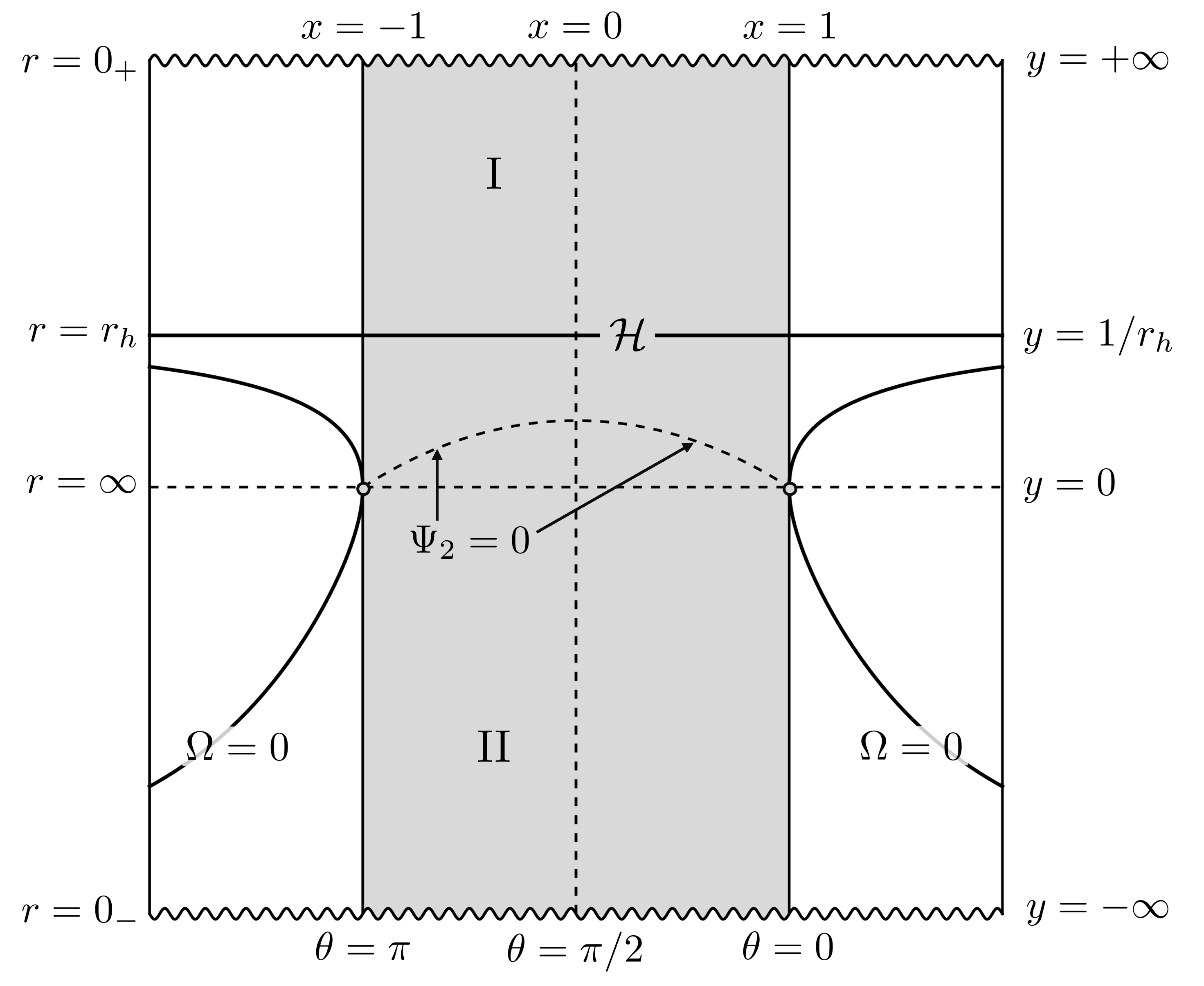}
    \caption{The structure of the Schwarzschild-BR spacetime in the $x-y$ coordinates. Particular parameters for this scheme are $m=1,~B=0.5$}
    \label{xy_scheme}
\end{figure}

In Fig. \ref{xy_scheme} we plot the corresponding curve on which $\Omega=0$ in the $x-y$ coordinates. As we have discussed above, indeed for $x\in (-1,1)$, there is no hypersurface for which geodesic particles require an infinite affine parameter to reach it, and thus the motion is only limited by the singularities $y=-\infty$ and $y=+\infty$. For $x=\pm 1$, $\Omega$ becomes zero for $y=0$, which means that geodesic particles, moving along the poles, will reach $r=\infty$ in the infinite proper time, meaning that the points $(x,y)=(1,0)$ and $(x,y)=(-1,0)$ represent the proper conformal boundary.

Also on Fig. \ref{xy_scheme} we plot the black hole horizon $\mathcal{H}$ that can be found by solving the equation $Q=0$, giving \textit{the only root} 
\begin{align}
    r_h=\dfrac{2m}{1+B^2\,m^2}.
\end{align}
This horizon separates two causally different regions I and II on Fig. \ref{xy_scheme}. 

Also, we plot the hypersurface at which $\Psi_2=0$, and by using (\ref{Psi_2_Schw_BR}) one can find that this happens at 
\begin{align}
    y=B^2 m (1-x^2).
\end{align}

At this surface, the spacetime becomes conformally flat; however, we do not consider this region as a boundary of the spacetime because the geodesic particles can easily cross this boundary.

By analyzing Fig. \ref{xy_scheme}, we thus can understand the main idea of how to continuously extend the Schwarzschild-BR spacetime so that it becomes geodesically maximal. For this, in the case $x\in (-1,1)$ we have to glue the coordinate patch $r\in (0,+\infty)$ with the coordinate patch $r\in (-\infty,0)$. This gluing is smooth because in the $y=1/r,~x=\cos\theta$ coordinates the metric (\ref{metr_y})-(\ref{Om_y}) is \textit{smooth} as one conducts the transition from $y=0_+$ to $y=0_-$. For the case of $x=\pm 1$ the situation is somewhat simpler, because in this case at $r\to+\infty$ the conformal factor tends to zero, and thus $r\to +\infty$ really represents a conformal boundary.

\begin{figure}
    \centering
    \includegraphics[width=1\linewidth]{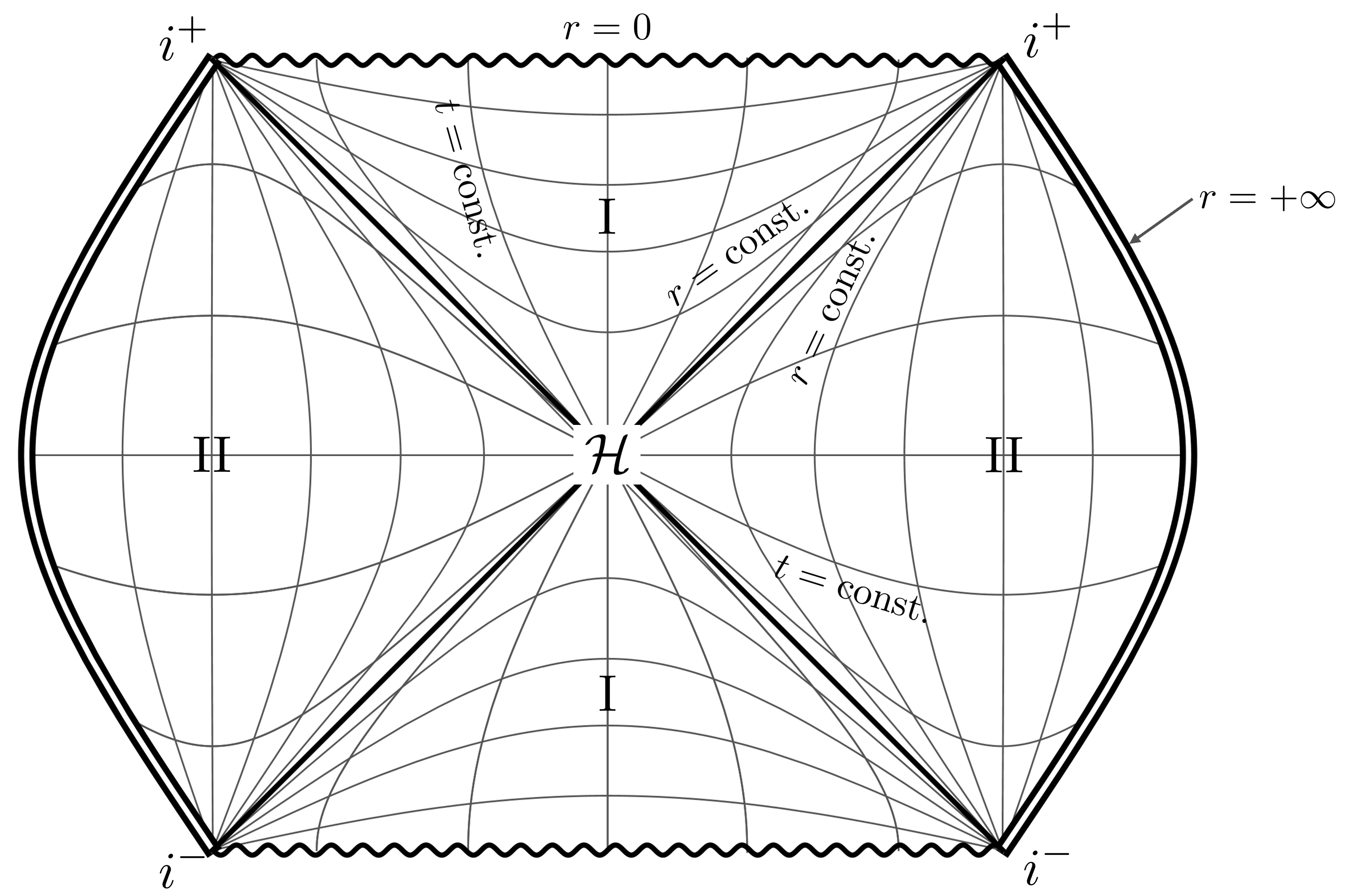}
    \caption{The Carter-Penrose diagram for the Schwarzschild-BR spacetime for $\theta=0,\pi$. The parameters chosen for this diagram: $m=1,~B=0.5$.}
    \label{fig_Schw_BR_Pen_2}
\end{figure}

We can physically interpret this continuation in such a way. One can see the region with negative $r$'s as the negative-mass black hole that is supported by the observation that the metric (\ref{Schw_BR_metr}) is invariant with respect to the change $r\to-r$ and $m\to -m$. However, the negative mass, instead of attraction, leads to the repulsion of particles. This means that any kind of particle that enters the $r<0$ region becomes repulsed from it. Moreover, if the particle is massive, it can never reach the singularity (for the negative-mass Schwarzschild, it is a known effect; see \cite{Bonnor1989}), is reflected, and returns to the $r>0$ patch. The only type of particles that can reach the singularity are the massless ones. This is reflected in the Penrose diagram Fig. \ref{fig_Schw_BR_Pen_2}; on this diagram, lightlike particles starting at any point of this diagram (except for the horizon) end at the singularity.

Thus, the magnetic field creates kind of ``confinement box''. What is interesting, this situation is analogous to the other model of magnetized black holes, found by Ernst and Wild \cite{Ernst1976,Ernst1976_2} that describe black holes in the Bonnor-Melvin background. Namely, in this spacetime the massive particles cannot escape to infinity while this possibility appears only for massless ones \cite{Galtsov1978}. This observation is interesting as it shows that spacetimes with different background electromagnetic fields may be common from the perspective of the global structure.

Now let us proceed with constructing the horizon-penetrating coordinates and drawing the Carter-Penrose diagram for the Schwarzschild-BR spacetime. 

For this we at first introduce null coordinates
\begin{align}
    u=t-r_*,~~~~~~v=t+r_*,
\end{align}
with the tortoise coordinate
\begin{align}
    r_*=\int\dfrac{\dd r}{\mathcal{Q}}.
\end{align}
In these coordinates, the metric (\ref{Schw_BR_metr}) becomes
\begin{align}
    \dd s^2=\dfrac{1}{\Omega^2}\Big[-\mathcal{Q}\,\dd u \dd v+r^2\Big(\dfrac{\dd \theta^2}{P}+P \sin^2\theta \dd \varphi^2\Big)\Big]
\end{align}

By using the explicit form of $\mathcal{Q}$, namely (\ref{Q_Schw_BR}), one obtains that the tortoise coordinate $r_*$ is given by
\begin{align}
    r_*^{\pm}=&k_h \ln \big|r(1+B^2m^2)-2m\big|\nonumber\\
    &-\dfrac{k_h}{2}\ln \big|(1-B^2 m r)^2+B^2r^2\big|\\
    &+k_o \arctan\big(B(r(1+B^2 m^2)-m )\big)+r_i^{\pm},\nonumber
\end{align}
where
\begin{align}
    k_h=\dfrac{2 m}{(1+B^2m^2)^2},~~~~~~k_o=\dfrac{1-B^2m^2}{B(1+B^2m^2)^2}.
\end{align}
Here the index $\pm$ is added to distinguish between the tortoise coordinates in the patches $r\in(0,+\infty)$ and $r\in(-\infty,0)$; $r_i^{\pm}$ are the corresponding integration constants in these patches. In principle, one could omit these integration constants, but it will be important to choose them properly for different regions we are gluing. For example, in the $r\in (0,+\infty)$ branch it is convenient to choose this constant in such a way that $r_*^+\to 0$ as we approach the singularity $r\to 0$. This can be done if we choose 
\begin{align}
    r_i^+=k_o\arctan(Bm)-k_h \ln(2m).
\end{align}

In the patch $r\in(-\infty,0)$ we choose this constant in such a way that the tortoise coordinate $r_*^+$ transfers from the region $r\in (0,+\infty)$ to the tortoise coordinate $r_*^-$ in the region $r\in (-\infty,0)$ in the connected way, namely
\begin{align}
    r_*^+\big|_{r=+\infty}=r_*^-\big|_{r=-\infty}.
\end{align}

This requirement leads to the unique choice of $r_i^-$ that is given by
\begin{align}
    r_i^-=k_o\pi+r_i^+.\label{223}
\end{align}

The next step is to introduce the Kruskal-Szekeres-like coordinates. We introduce them in such a way.
\begin{align}
    &\mathrm{In~the~patch~}r\in(0,+\infty):\nonumber\\
    &\begin{cases}
         U=\dfrac{1}{2}\Big[\exp\Big(\dfrac{t+r_*^+}{2k_h}\Big)-\epsilon_+\exp\Big(-\dfrac{t-r_*^+}{2k_h}\Big)\Big] , \\
          V=\dfrac{1}{2}\Big[\exp\Big(\dfrac{t+r_*^+}{2k_h}\Big)+\epsilon_+\exp\Big(-\dfrac{t-r_*^+}{2k_h}\Big)\Big].
    \end{cases}\label{UV_def_1}\\
    &\mathrm{In~the~patch~}r\in(-\infty,0):\nonumber\\&\begin{cases}
         U=\dfrac{1}{2}\Big[\exp\Big(\dfrac{t+r_*^-}{2k_h}\Big)-\epsilon_-\exp\Big(-\dfrac{t-r_*^-}{2k_h}\Big)\Big],  \\
          V=\dfrac{1}{2}\Big[\exp\Big(\dfrac{t+r_*^-}{2k_h}\Big)+\epsilon_-\exp\Big(-\dfrac{t-r_*^-}{2k_h}\Big)\Big],
    \end{cases}\label{UV_def_2}
\end{align}
where 
\begin{align}
    \epsilon_{\pm}=\mathrm{sign}(r)\,\mathrm{sign}\big(r(1+B^2m^2)-2m\big).
\end{align}

Here $\pm$ stands for the sign of $r$. In (\ref{UV_def_1})-(\ref{UV_def_2}) different $\epsilon$ quantities are used, depending on the sign of $r$.

In these coordinates, the metric is regular when crossing through the horizon, and also these coordinates are continuous when transferring from the $r\in(0,+\infty)$ to $r\in (-\infty,0)$ patch. To visualise these coordinates and draw the Carter-Penrose diagrams, we compactify these coordinates by using the standard relations.
\begin{align}
    \eta=&\arctan(V+U)+\arctan(V-U),\\
    \chi=&\arctan(V+U)-\arctan(V-U).
\end{align}

In these coordinates, we can plot the Carter-Penrose diagram.

First of all, we plot it for the $\theta=\{0,\pi\}$. In this case, as we mentioned in the discussion of Fig. \ref{xy_scheme}, the conformal factor tends to zero as $r\to +\infty$. The geodesic observers reach this point in an infinite proper time, so the surface $r\to +\infty$ is the proper conformal boundary of the Schwarzschild-BR spacetime in this case.

The resulting Carter-Penrose diagram is plotted in Fig. \ref{fig_Schw_BR_Pen_2}. This spacetime has a curvature singularity at $r=0$, and if we track the observers with $t=\mathrm{const}.$, one can see that such observers start at the singularity $r=0$, which is the same as for the Schwarzschild spacetime. Then, as $r$ increases, such an observer crosses the horizon $\mathcal{H}$. After the horizon, such an observer reaches the conformal boundary $\mathcal{I}$ at $r=+\infty$.

The observers with $r=\mathrm{const}.$ end at the $i^+$ spacelike infinity for large positive $t$. For large negative $t$ they reach the $i^-$ spacelike infinity. 

\begin{figure}
    \centering
    \includegraphics[width=1\linewidth]{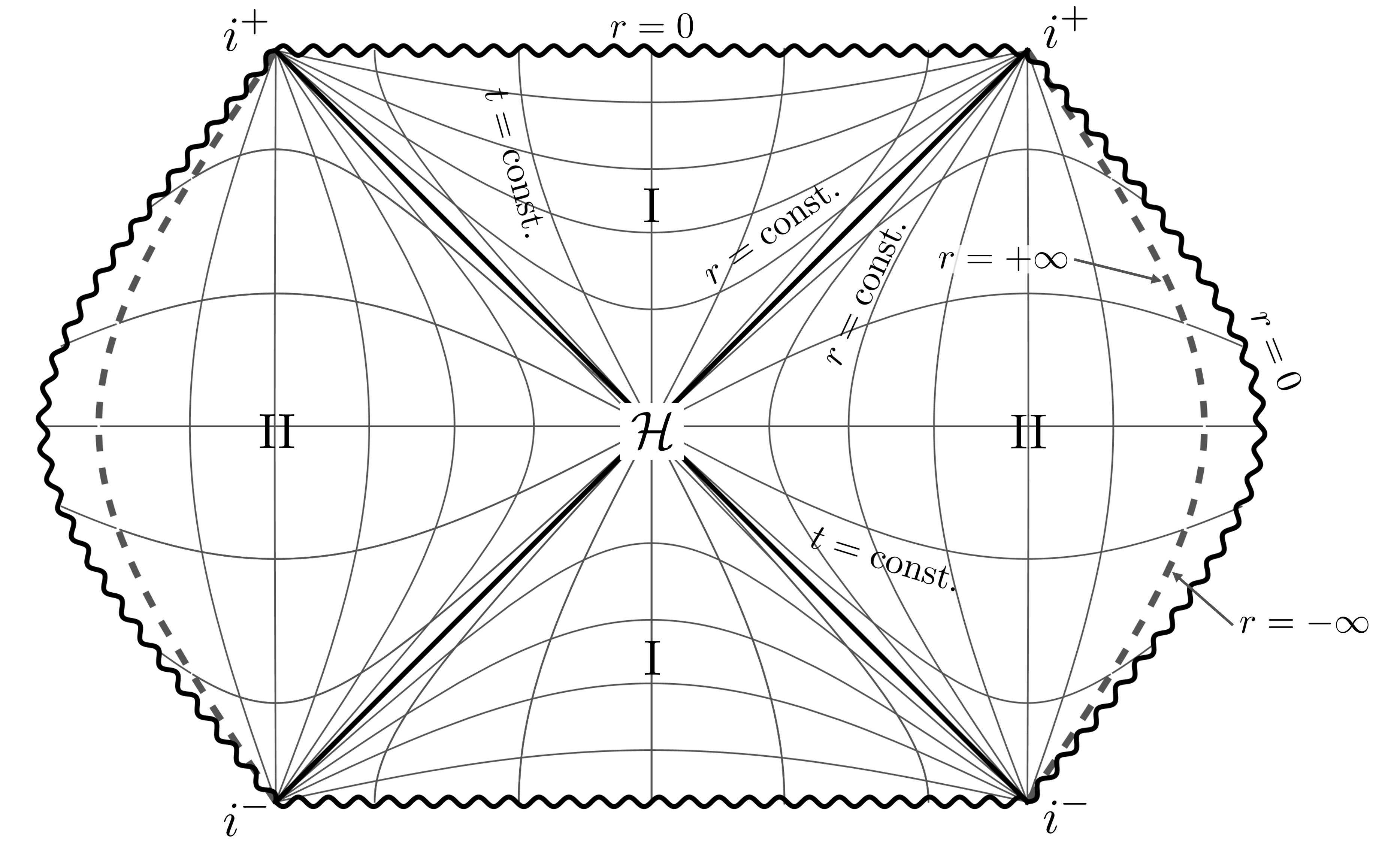}
    \caption{The Carter-Penrose diagram for the Schwarzschild-BR spacetime for $\theta\neq 0,\pi$. The parameters chosen for this diagram: $m=1,~B=0.5$.}
    \label{fig_Schw_BR_Pen_1}
\end{figure}

Now let us investigate the second case, corresponding to $\theta\notin \{0, \pi\}$, shown in Fig. \ref{fig_Schw_BR_Pen_1}. For $r\in (0,+\infty)$, the structure of the spacetime is the same as in Fig. \ref{fig_Schw_BR_Pen_2}. However, after the horizon, an observer with $t=\mathrm{const}.$ reaches the hypersurface $r=+\infty$ that is continuously transferred to $r=-\infty$ (dashed curve in Fig. \ref{fig_Schw_BR_Pen_1}). Then such an observer reaches $r=0$ from the directions of the negative values. 

The observers with $r=\mathrm{const}.$ end at $i^+$ (if time is increasing) or at $i^-$ (if time is decreasing).

\begin{figure*}
    \includegraphics[width=0.9\textwidth]{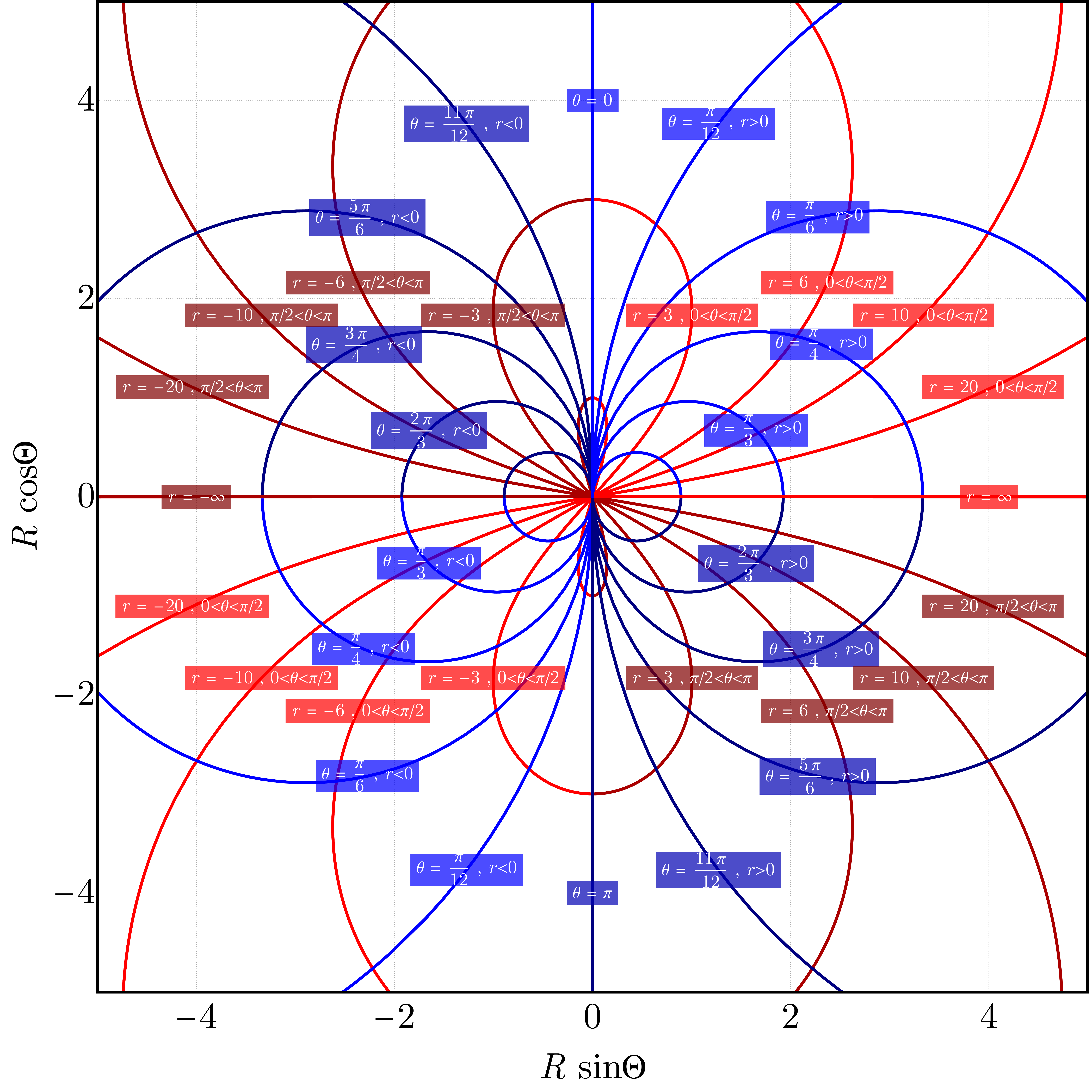}
    \caption{The lines of constant $r$ (red curves) and $\theta$ (blue curves) in the Bertotti--Robinson spherical coordinates $R$ and $\Theta$. Light blue and light red colors correspond to the $\theta\in [0,\pi/2]$, dark blue and red colors correspond to $\theta\in [\pi/2,\pi]$.}
    \label{fig_6}
\end{figure*}

Now let us explain the reason why the conformal structure is so unusual. For this, let us consider the limit of zero mass of a Schwarzschild-BR black hole, and inspect the lines of constant $r$ and $\theta$ coordinates. In the massless case, the Bertotti--Robinson spacetime can be obtained by the transformation of coordinates (2.16) from Sec. II.B in \cite{Podolsky2025}. Here we repeat this transformation (notice that, as we are dealing with the non-twisting spacetime, we set $a=0$) 
\begin{align}
    \dfrac{R^2}{e^2}=\dfrac{1+B^2r^2}{\Omega^2}-1,~~~e \sin \Theta=\dfrac{r}{\Omega}\sin \theta,
\end{align}
where $e=B^{-1}$, $\Omega^2=1+B^2r^2 \sin^2\theta$, the rest of the coordinates remain the same.
In Fig. \ref{fig_6} we plot the lines of constant $r$ and $\theta$ in the plane $R\sin\Theta$ and $R\cos\Theta$.

Let us inspect the corresponding curves of constant $r$ and $\theta$ in Fig. \ref{fig_6}. We fix $\theta$ to some constant and observe what changes as $r$ goes from zero to infinity (for definiteness, let us take some $\theta\in [0,\pi/2]$, which corresponds to the light blue curves in Fig. \ref{fig_6}). For zero $r$, the coordinate $R$ is also zero, so the origins in these two coordinates coincide. Then, as one increases $r$, the distance to the origin also increases, but as $r\to +\infty$, the coordinate $R$ does not generally reach infinity. $R$ goes to infinity only for $\theta=0$ or $\theta=\pi$, for all other values of $\theta$, $R$ \textit{remains finite}. Then there is an interesting point. If one performs the transformation $\varphi\to\varphi+\pi$, then this curve smoothly transforms into a part corresponding to $r\to-r$. When one further goes along this curve and increases $r$ from $-\infty$ to $0$, the coordinate $R$ thus decreases down to zero. Therefore, in the BR background coordinates, one can clearly see that by ``gluing'' the two parts of spacetime at $r\to+\infty$ and $r\to-\infty$, one obtains a \textit{closed curve} that returns to $R=0$.\footnote{Notice that we have to also change $\varphi\to\varphi+\pi$: but this need not be an issue, as for the Schwarzschild-BR spacetime $\partial_{\varphi}$ is the Killing vector. The same is true for the whole family of Kerr-Newman-BR spacetimes.} However, for $\theta\in\{0,\pi\}$ one can easily reach $R\to\infty$. 

This result is quite interesting, as it allows us to understand what happens when we add a black hole in the spherical coordinates, adapted to the Bertotti--Robinson background. If we add a mass $m$ at the center of Fig. \ref{fig_6}, the coordinate lines of constant $\theta\notin \{0,\pi\}$ still have the property that the $R$ coordinate is finite as $r\to \infty$. Because of that, one cannot reach the conformal boundary of this spacetime that is located at $R\to \infty$, and when reaching $r\to+\infty$ one simultaneously switches to the curve with negative $r$'s and returns to the origin. This explains the conformal diagram Fig. \ref{fig_Schw_BR_Pen_1}\footnote{We want to notice that even in the recent work \cite{Zhou2026_3}, where authors found BR-like coordinates for the general Kerr-Newman-Bertotti-Robinson spacetime, this problem is not fully resolved as their $R\to \infty$ limit corresponds to $r\to \infty$ with $\theta\to 0$ or $\theta\to \pi$}. 

However, if $\theta\in\{0,\pi\}$, then as $r$ increases to $+\infty$, the $R$ \textit{also increases to} $+\infty$ and such an observer sees the conformal boundary there. This supports the observations made while discussing Fig. \ref{fig_Schw_BR_Pen_2}.

\section{Conformal structure of generic Kerr-Newman-BR spacetime}

Now, let us proceed with the analysis of the conformal structure of the most general Kerr-Newman-BR spacetime. In this case, we use the metric presented in (2.1)-(2.5) in \cite{Ovcharenko2026_2}, namely
\begin{align}
\dd s^2= \dfrac{1}{\Omega^2}\Big[&
    -\dfrac{Q}{\rho^2}\big(\dd t-a \sin^2\theta\, \dd \varphi\big)^2
    +\dfrac{\rho^2}{Q}\,\dd r^2 + \dfrac{\rho^2}{P}\,\dd\theta^2\nonumber\\
    &+\dfrac{P}{\rho^2} \sin^2\theta \big(a\dd t-(r^2+a^2)\dd \varphi\big)^2\Big],\label{metr}
\end{align}
where
\begin{align}
    \rho^2 &= r^2+a^2\cos^2\theta \,, \label{rho_eq}\\
    P &= 1 + B^2 \mu^2 \cos^2\theta\,, \label{P_eq}\\
    Q &= I\,\Delta\,, \label{Q_eq}\\
    \Omega^2 &= I-B^2 \Delta \cos^2\theta\,, \label{Om2_eq}\\
    I &= (1 +k\,B\,r)^2+B^2 r^2\,, \label{I_eq}\\
    \Delta   &= (1+k^2)\,a^2 -2m\,r + (1 + k^2 - e^2/a^2)\, r^2\,, \label{Delta_eq}
\end{align}
with the dimensionless constants
\begin{align}
    k   &= \dfrac{e\, s+a\, m\, B}{a\,(1+e^2 B^2)}\,,\label{c_eq}\\
    s^2 &= 1+(e^2-a^2)B^2-(m^2+e^2)\,a^2 B^4\,,\label{s_eq}\\
    \mu^2&=m^2-(1+k^2)a^2+(1+k^2)e^2.
\end{align}
As in the previous section, the $r\to +\infty$ limit does not vanish the Weyl tensor (unless the magnetic field is zero, see Sec. III.A. in \cite{Ovcharenko2026_2}). As in the case $a=0=e$, we must extend the domain of the coordinates used so that we obtain a geodesically maximal manifold. This requires finding physical singularities and the surfaces where the affine parameter of geodesical particles becomes infinite.

For this, we, as in the previous section, consider such a transformation of the coordinates
\begin{align}
    x=\cos\theta,~~~y=\dfrac{1}{r}.
\end{align}

\begin{figure}
    \centering
    \includegraphics[width=1\linewidth]{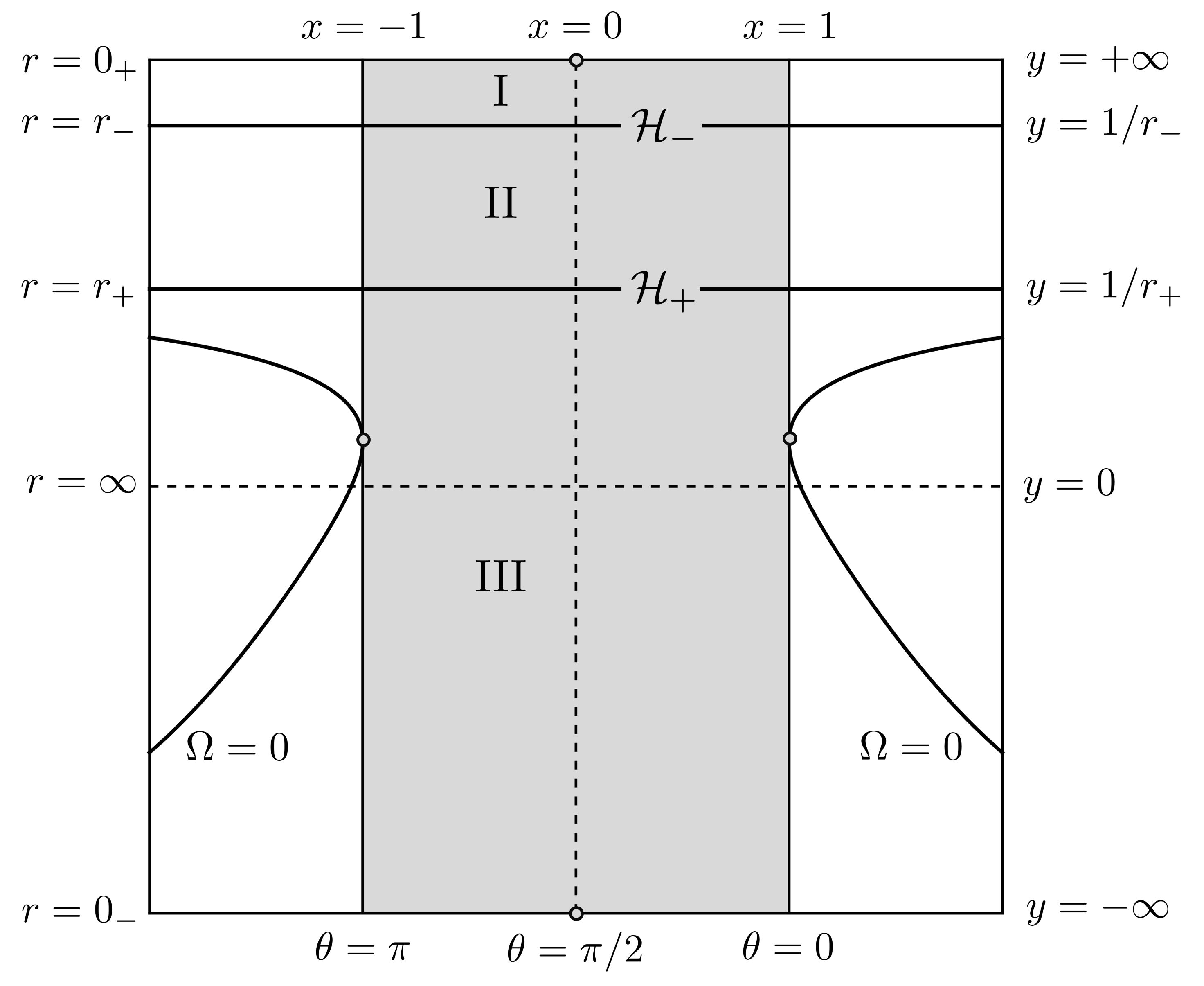}
    \caption{Structure of the Kerr-Newman-BR spacetime in the $x-y$ coordinates. Parameters used for this scheme: $m=1,~e=0.2,~a=0.5,~B=-0.4$}
    \label{xy_scheme_2}
\end{figure}

In Fig. \ref{xy_scheme_2}, we show the structure of the Kerr-Newman-BR spacetime in these coordinates for a set of parameters when there are two distinct horizons $\mathcal{H}_-$ and $\mathcal{H}_+$. One can see that there are several significant differences with the scheme in Fig. \ref{xy_scheme}. First of all, we notice that the structure of the singularities is different. In the general case (when $a\neq 0$), the singularity is placed only at $r=0$ and $\theta=\pi/2$ (see Sec. IV.A in \cite{Ovcharenko2026_2}). In Fig. \ref{xy_scheme_2}, these ``ring'' singularities are presented as gray points at $r=0$, $\theta=\pi/2$.
Let us discuss other differences between Fig.~\ref{xy_scheme} and Fig.~\ref{xy_scheme_2}. Fig~\ref{xy_scheme_2} has two horizons separating the coordinate patches I, II, and III, while Fig. \ref{xy_scheme} contains only one horizon $\mathcal{H}$. Also, there is no surface at which $\Psi_2$ becomes zero (unlike the case shown in Fig. \ref{xy_scheme}). Even though such a surface is not of great importance for us, it is interesting that when one adds a rotation, $\Psi_2$ may become zero only when $\Omega=0$.

The surfaces at which the affine parameter is infinite are given by points at which $\Omega^2=0$. In Fig. \ref{xy_scheme_2}, these points are shown as thick curves. As in the previous subsection, they touch the realistic range of coordinate $x\in [-1,1]$ \textit{only} at $x=-1$ or $x=+1$ (these points are represented as gray dots). This is the same property as the one observed in Fig. \ref{xy_scheme}, but the position of the gray dots is shifted. Also, we see that the Kerr-Newman-BR spacetime in the $(x,y)$ coordinates can be easily analytically continued through the surface $r\to+\infty$ by gluing it with the coordinate patch $r\to -\infty$.\footnote{The fact that this gluing is smooth follows from the fact that this spacetime is smooth in $(x,y)$ coordinates, where the transfer $y=0_-$ to $y=0_+$ is smooth.}

With this scheme, we can better understand the structure of the spacetime. Thus, if we consider an observer that is moving in the equatorial plane ($x=0,~\theta=\pi/2$), such an observer starts at the singularity at $r=0$, crosses the inner $\mathcal{H}_-$ and outer $\mathcal{H}_+$ horizons, and reaches $r\to+\infty$. Then, as this patch is glued with $r\to -\infty$, the observer reaches $r=0_-$ from the side of negative values.

If the observer moves along the poles ($x=1,~\theta=0$ or $x=-1,~\theta=\pi$), then the situation is somehow different. If such an observer starts at $r=0$ and moves in the direction of an increasing $r$, then it crosses the inner and outer horizons and reaches the proper conformal infinity where $\Omega^2=0$. If the same observer starts its motion from $r=0$ and moves in the direction of negative $r$'s (which is possible because in the case of motion along the poles, the observer does not see a singularity at $r=0$), it ends up in the lower part of the scheme and reaches the conformal infinity from the other side.

However, there appears to be a question of what happens when the observer is not moving either in the equatorial plane nor along the poles ($x\notin\{-1,0,1\}$)? In this case, as one can see from the scheme Fig. \ref{xy_scheme_2}, there are no singularities and conformal infinities. This means that there is no surface at which such a manifold can end. However, as we wish to make such a manifold geodesically maximal, we have to resolve this issue somehow, and we propose to identify the points $r=0_+$ and $r=0_-$ in Fig. \ref{xy_scheme_2}. If one does so, the observer at $x\notin\{-1,0,1\}$ that starts at $r=0_+$ and increases its $r$ crosses the inner and outer horizons, crosses the surface of $r\to +\infty$, and ends up in the coordinate patch with negative $r$, then increases its $r$ and ends up at $r=0_-$, which is identified with $r=0_+$, and the situation repeats. Also, we wish to mention that because of this identification, the patches I and III are identical. 

The interpretation of this identification is not so straightforward, because after the return to $r=0$, it is not clear whether we return to the same black hole spacetime, or we end up in a ``parallel universe''. To avoid possible issues with causality, we will interpret such a spacetime as a parallel one. Thus, one obtains a wormhole-like spacetime, as reported in \cite{Zhou2026}.

Now, let us move to explicitly constructing the horizon-penetrating coordinates and drawing the Carter-Penrose diagram for the Kerr-Newman-BR spacetime. 

\begin{samepage}
For this, we at first introduce null coordinates
\begin{align}
    u=t-r_{*},~~~v=t+r_{*},
\end{align}
with the tortoise coordinate
\begin{align}
    r_{*}=\int \dfrac{r^2+a^2}{Q}\dd r.\label{313}
\end{align}
\end{samepage}

The reason why these coordinates are horizon-penetrating was proven in Sec. III.D. in \cite{Vratny2021} for the subcase $l=0$. The reason why this analogy can be used is that the metric (\ref{metr}) from this work and the metric (1) in \cite{Vratny2021} in the $l=0$ case are the same \textit{off-shell}. This means that the same algorithms can be used, but the explicit expressions and the conformal structure are different.

By integrating (\ref{313}), we obtain such an expression for the tortoise coordinate
\begin{align}
    r_{*}^{\pm}=&k_{h}^+\ln|r-r_+|+k_{h}^-\ln|r-r_-|\\&+k_o\arctan\big(B(1+k^2)r+k\big)+k_l \ln (I(r))+r_i^{\pm},\nonumber
\end{align}
where
\begin{widetext}
\begin{align}
    &k_h^+=\dfrac{a^2+r_+^2}{(1+k^2-e^2/a^2)(r_+-r_-)I(r_+)},\nonumber\\
    &k_h^-=\dfrac{a^2+r_-^2}{(1+k^2-e^2/a^2)(r_--r_+)I(r_-)},\nonumber\\
    &k_o=\dfrac{1-B^2(1-k^2)r_- r_++B k(r_-+r_+)-a^2 B^2\Big(1-\big(k+B(1+k^2)r_-\big)\big(k+B(1+k^2)r_+\big)\Big)}{B(1+k^2-e^2/a^2) \,I(r_-)I(r_+)},\\
    &k_l=-\dfrac{(r_-+r_+)\big(1-a^2B^2(1+k^2)\big)+2 B k(r_-r_+-a^2)}{2(1+k^2-e^2/a^2)I(r_-)I(r_+)},\nonumber
\end{align}
\end{widetext}
where $I(r)$ is the function, given by (\ref{I_eq}), $r_-$ and $r_+$ are the inner and outer horizons respectively. The constants $r_i^{\pm}$ are the integration constants that have to be chosen distinctly at various coordinate patches. In the patch $r\in (0,+\infty)$, we choose this constant in such a way that $r_{*}^+(r=0)=0$. This becomes possible when we choose
\begin{align}
    r_i^+=-k_h^+ \ln r_+-k_h^- \ln r_--k_o\arctan k .
\end{align}
In the patch $r\in (-\infty,0)$ we do the same choice as we did in the previous section, namely the conditions 
\begin{align}
    r_{*}^+\big|_{r\to +\infty}=r_{*}^-\big|_{r\to -\infty}.
\end{align}
From this condition, we obtain 
\begin{align}
    r_i^-=k_o\pi+r_i^+,
\end{align}
the same as (\ref{223}).

Thus we constructed the globally smooth tortoise coordinates. The Kruskal-Szekeres-like coordinates can be constructed knowing them in a standard way (we do not present details of this construction; they can be found in \cite{supp_mat}). The only thing that has to be noticed is that we have to introduce one set of coordinates that smoothly goes through the inner horizon, and another set of coordinates that smoothly goes through the outer horizon. Below, we present only the Carter-Penrose diagrams for various cases. 

\begin{figure*}
    \centering
    \includegraphics[width=0.9\textwidth]{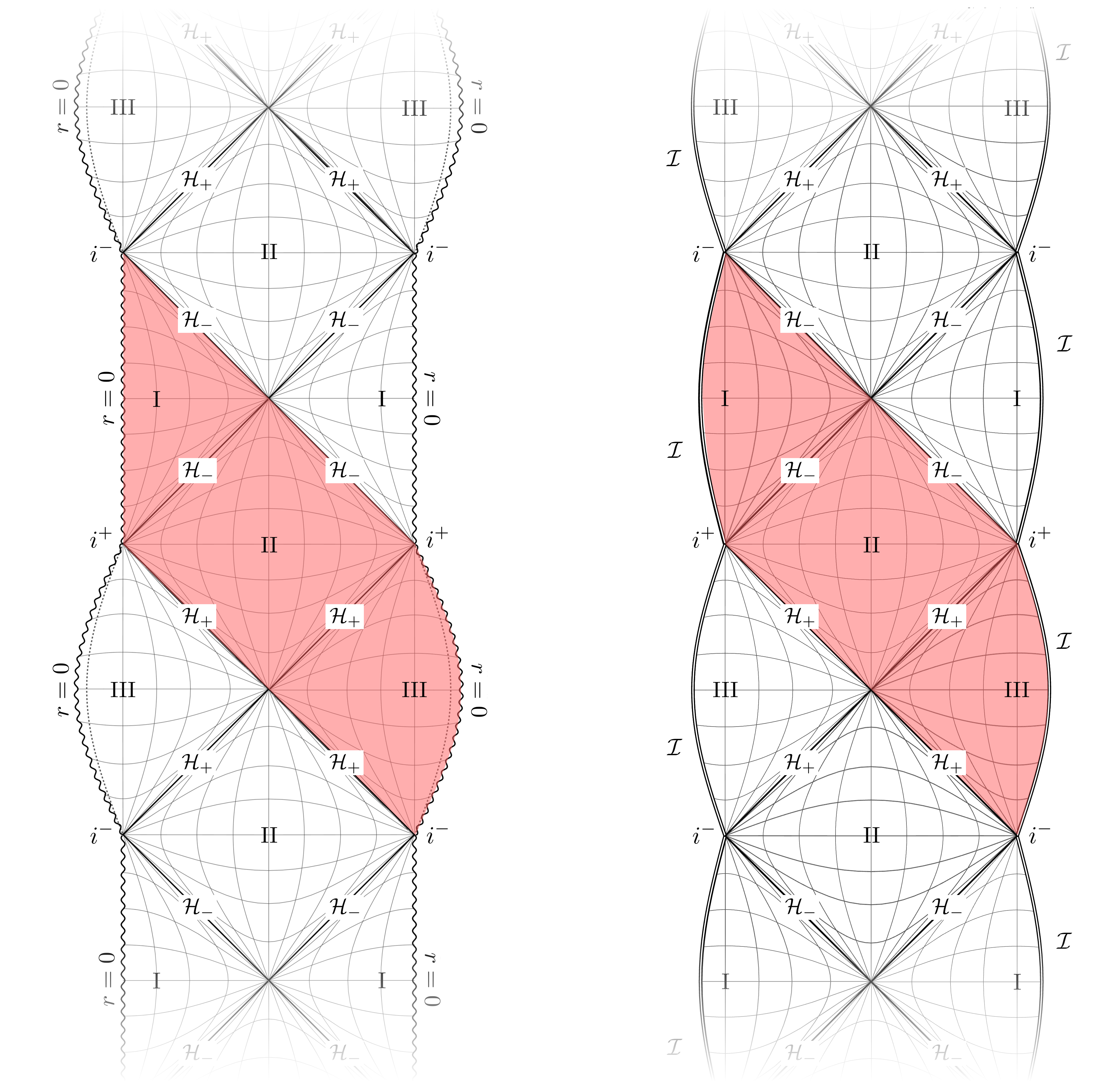}
    \caption{The Carter-Penrose diagram for the Kerr-Newman-BR spacetime for $\theta=\pi/2$ (left part) and $\theta=0,\pi$ (right part). The parameters used for this diagram: $m=1,~e=0.2,~a=0.7,~B=-0.4$}
    \label{fig_KN_BR_Pen_1}
\end{figure*}

The corresponding Carter-Penrose diagrams for the case of the equatorial plane ($\theta=\pi/2$) and poles ($\theta=0,\pi$) are given in Fig. \ref{fig_KN_BR_Pen_1}. On these diagrams, we see all the elements we have discussed so far when investigating the diagram in Fig. \ref{xy_scheme_2}. Namely, if we restrict the observer to the equatorial plane, then the constant $t$ observers at first cross the inner horizon, then cross the outer horizon, then reach $r\to+\infty$ (the dashed curve), appear in the patch with negative $r$ and reaches the singularity at $r=0$ from the other side. The observers with the constant $r$ reach either $i^+$ (if the time coordinate is increasing) or $i^-$ (if the time coordinate is decreasing). For the second case, when we restrict an observer to the poles $\theta\in\{0,\pi\}$, the observer with constant $t$ reaches the conformal boundary from either one side (if $r$ is increasing) or from another side (if $r$ is decreasing). The observers with constant $r$ end either on $i^+$ (if the time coordinate is increasing) or on $i^-$ (if the time coordinate is decreasing).

\begin{figure*}
    \centering
    \includegraphics[width=1\textwidth]{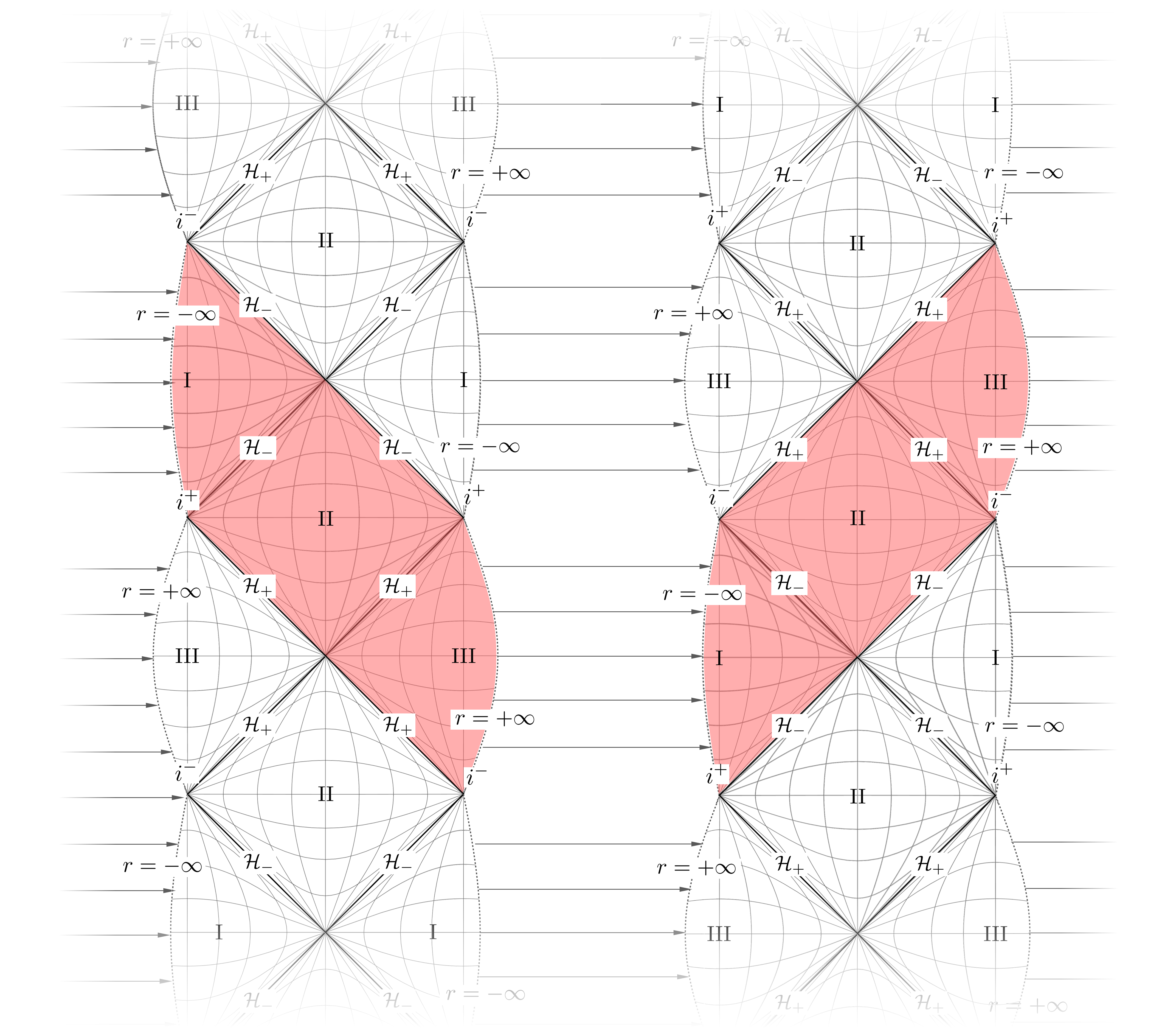}
    \caption{The Carter-Penrose diagram for the Kerr-Newman-BR spacetime for $\theta\neq0,\pi/2,\pi$. the parameters used for this diagram: $m=1,~e=0.2,~a=0.7,~B=-0.4$}
    \label{fig_KN_BR_Pen_2}
\end{figure*}

We also show an analogous diagram for the case of $\theta\notin\{0,\pi/2,\pi\}$, see Fig. \ref{fig_KN_BR_Pen_2}. In this case, as we mentioned, we have to ``glue'' the different coordinate patches from the parallel diagrams. Thus, the observer, when reaching $r\to +\infty$, moves to the ``parallel universe'' starting at $r\to-\infty$. One can try to glue the $r>0$ and $r<0$ patches of the same universe; however, in such a case, while crossing the boundary $r=\infty$ one would go from the patch III of the left side of the diagram Fig. \ref{fig_KN_BR_Pen_2} to the patch I of the same left side of the diagram Fig. \ref{fig_KN_BR_Pen_2}. However, these patches have different Penrose time $\eta$. This would lead to time travel after crossing $r=\infty$.

\begin{figure*}
    \centering
    \includegraphics[width=0.9\textwidth]{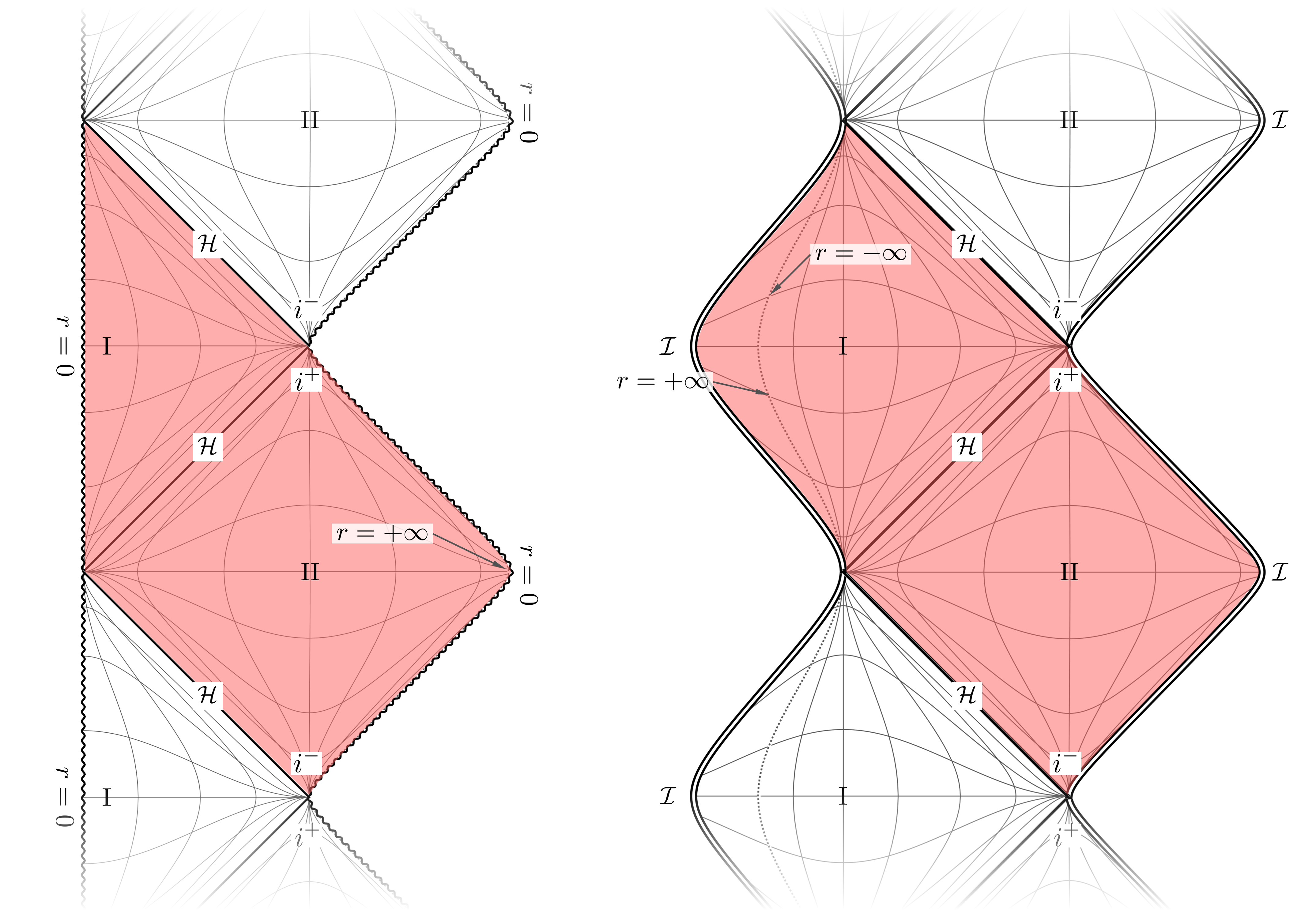}
    \caption{The Carter-Penrose diagram for extremal Kerr-Newman-BR spacetime for $\theta=\pi/2$ (left part) and $\theta=0,\pi$ (right part). The parameters used for this diagram: $m=1,~a=0.5,~B=-0.4$. The charge $e$ is chosen in such a way that the spacetime is extremal; a particular value of charge is $e\approx 0.3984953100244539$.}
    \label{fig_KN_BR_Pen_extr_1}
\end{figure*}

\begin{figure*}
    \centering
    \includegraphics[width=0.9\textwidth]{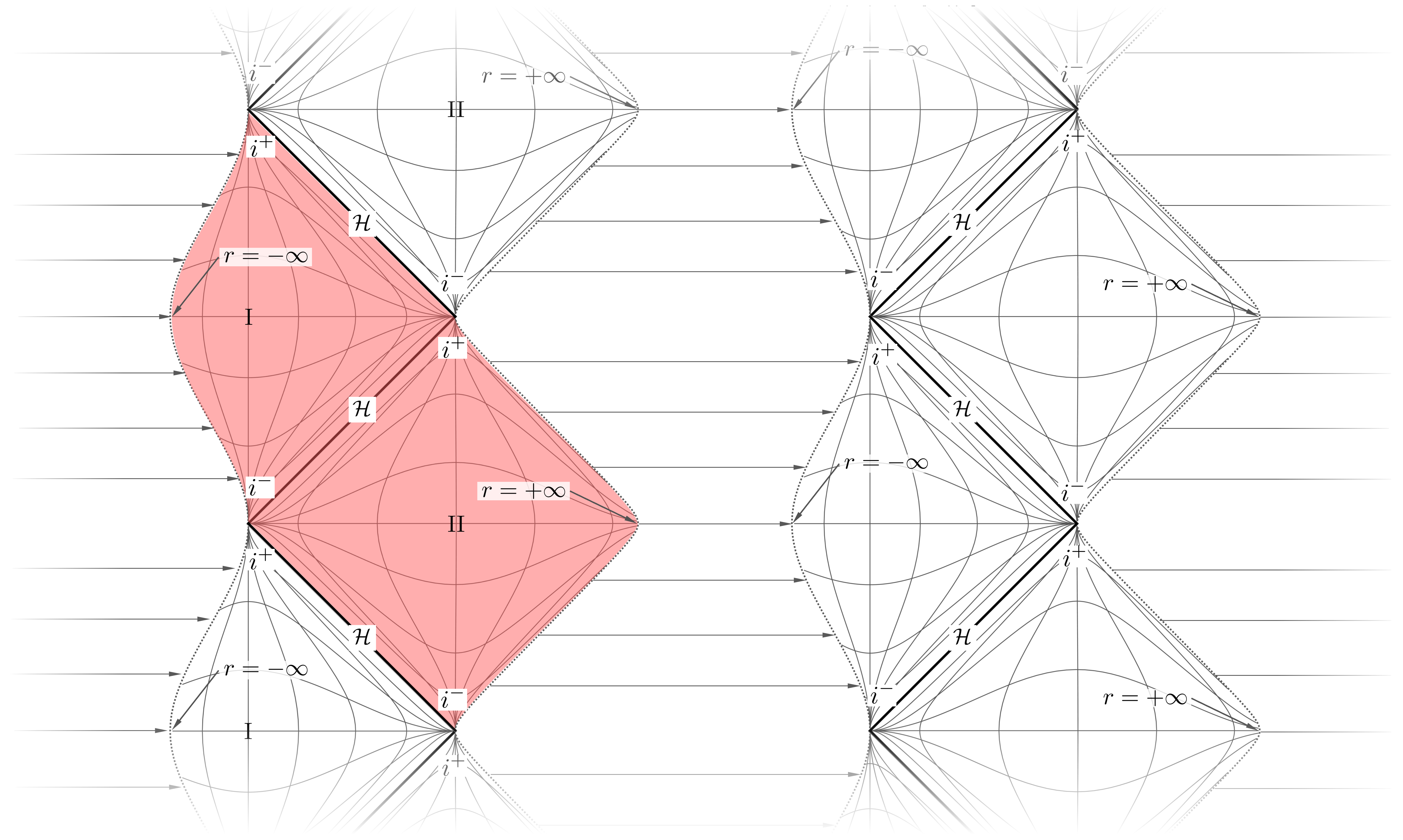}
    \caption{The Carter-Penrose diagram for extremal Kerr-Newman-BR spacetime for $\theta\neq0,~\pi/2,~\pi$. The parameters used for this diagram: $m=1,~a=0.5,~B=-0.4$. The charge $e$ is chosen in such a way that the spacetime is extremal; a particular value of charge is $e\approx 0.3984953100244539$.}
    \label{fig_KN_BR_Pen_extr_2}
\end{figure*}

At the end, we wish to analyze the case of \textit{extremal} black holes. Without explanation of technical details (that can be found in \cite{supp_mat}), we present only the result in Fig. \ref{fig_KN_BR_Pen_extr_1} and Fig. \ref{fig_KN_BR_Pen_extr_2}.

On these diagrams, we see the expected structure, namely that for the equatorial plane (left picture of Fig. \ref{fig_KN_BR_Pen_extr_1}), a spacelike observer starting near the singularity crosses the horizon, then reaches $r\to+\infty$, moves to the region with negative $r$, and then reaches $r=0$ from the side of negative values. Notice that the region between $r\to -\infty$ and $r=0$ is practically indistinguishable from the second $r=0$ surface. Near the poles (right picture of Fig. \ref{fig_KN_BR_Pen_extr_1}) an observer is bounded by conformal boundary $\mathcal{I}$, while if $\theta\notin\{0,\pi/2,\pi\}$ (Fig. \ref{fig_KN_BR_Pen_extr_2}), one also obtains ``parallel universe'' that, however, obtains only one horizon.

\section{Conclusions}

In this work, we were able to continuously extend the Schwarzschild-BR and the Kerr-Newman-BR spacetimes, creating a geodesically maximal manifold. We have found that for the Schwarzschild-BR spacetime, if $\theta\notin\{0,\pi\}$, there is a way to glue coordinate patches with positive and negative $r$ and obtain a geodesically maximal manifold ending with the $r=0$ singularities only, see Fig. \ref{fig_Schw_BR_Pen_1}. If $\theta\in\{0,\pi\}$, the spacetime ends with the conformal boundary at $\Omega=0$, see Fig. \ref{fig_Schw_BR_Pen_2}. For the case of the Kerr-Newman-BR spacetime, the situation is more complicated. In the equatorial plane $\theta=\pi/2$, as there is a singularity at $r=0$, the gluing of patches with positive and negative $r$ creates a geodesically maximal manifold that ends with the singularity at $r=0$, see the left part of Fig. \ref{fig_KN_BR_Pen_1}. For the polar directions $\theta\in\{0,\pi\}$, there is a conformal boundary at large $r$, but as $r=0$ is not a singularity for these directions, one can easily cross $r=0$ and end up on the other side of the conformal boundary, see the right side of Fig. \ref{fig_KN_BR_Pen_1}. The most interesting situation happens for $\theta\notin\{0,\pi/2,\pi\}$. In this case, by gluing the patches with positive and negative $r$, one obtains a manifold that does not have either the singularity (as $r=0$ is no longer the singularity) or the conformal infinity. This creates the situation depicted in Fig. \ref{fig_KN_BR_Pen_2} that represents an infinite combination of parallel black hole spacetimes that are connected by the ``throat''  at $r\to +\infty$. At the end, we also considered the case of extremal black holes (Fig. \ref{fig_KN_BR_Pen_extr_1} and Fig. \ref{fig_KN_BR_Pen_extr_2}) that have only one horizon. Except that all other global properties remain the same.

As was mentioned in \cite{Zhou2026}, this may be interpreted rather as a wormhole than as a black hole spacetime. Even though technically this interpretation is correct (and is supported by the calculations in this work and in \cite{Zhou2026,Zhou2026_2}), we still want to warn that this analytical extension was obtained by a specific ``gluing'' of the part of spacetime with positive $r$ with the identical copy but with negative $r$. Even though formally this gluing is correct (and does not produce any matter content at the gluing surface), this does not need to be \textit{the only} possible continuous extension of the Kerr-Newman-BR spacetime. In principle, there may exist an alternative way to extend the spacetime across $r\to +\infty$ that ends with the conformal boundary and does not lead to an infinite number of parallel universes. 

At the end, we would like to discuss the applicability of the cosmic conjecture to this spacetime. In its weak form, it is not violated as neither the Schwarzschild-BR nor Kerr-Newman-BR spacetimes contain future null infinity. However, in its strong form, it appears to be violated, as in both these cases, one deals with the timelike singularity, and any radiation coming from singularities may affect the evolution of the spacetime. The corresponding issue, however, also exists for the Kerr and Reissner-Nordstrom spacetimes, where the singularity is timelike and may affect the evolution of the spacetime. The significant difference is that this spacetime does not have a null infinity and is ``confined'' between two singularities, both of which can affect the spacetime evolution. Hyperbolicity can be restored only if we assume that no radiation is coming from the singularities.

\section{Acknowledgments}

This work has been supported by the Czech Science
Foundation Grant No. GA\v{C}R 26-22381S, and by the
Charles University Grant No. GAUK 260325. The author thanks Ji\v{r}\'{i} Podolsk\'{y} and David Kubiz\v{n}\'{a}k for valuable discussions.

\section{Data availability}
The data that support the findings of this article are openly available \cite{supp_mat}.

\appendix

\section{Motion of geodesic particles}\label{append}

In this Appendix, we describe the motion of geodesic particles in Kerr-Newman-BR spacetime and show that the affine parameter diverges near the surface $\Omega=0$. As in the used coordinates the singularity is achievable only at $\theta=0$ or $\theta=\pi$. Let us consider only the case with $\theta=0$ as $\theta=\pi$ is complementary. For this let us consider the Hamilton-Jacobi equation 
\begin{align}
	g^{\mu\nu}\dfrac{\partial S}{\partial x^2}=-\mu^2
\end{align} 
where $\mu$ is the mass of a particle. As KN-BR spacetime allows for separation of the variables, we can write
\begin{align}
	S=-E t+L\varphi +s(r,\theta)
\end{align}
where $E$ is the particle's energy and $L$ it's angular momentum. As we consider the motion along $\theta=0$ pole, we have to take $L=0$ and assume that $s$ is the function of $r$ only. By substituting this to the Hamilton-Jacobi equation, one obtains
\begin{align}
	s'^2=\dfrac{\rho_0^4}{Q^2\Omega_0^2}\Big(E^2\Omega^2-m^2\dfrac{Q}{\rho_0^2}\Big)
\end{align}
where $\rho^2_0=\rho^2(\theta=0)$, $\Omega^2_0=\Omega^2(\theta=0)$. The radial velocity is given by
\begin{align}
	m^2 \Big(\dfrac{d r}{d\lambda}\Big)^2=\Omega_0^2\Big(E^2\Omega_0^2-\dfrac{m^2 Q}{\rho_0^2}\Big).
\end{align}

What is interesting is that as $\Omega$ tends to zero, the RHS becomes negative for massive particles, thus giving complex radial velocity. This means that massive particles cannot reach the $\Omega=0$ boundary at all.

\end{document}